\documentclass{article}

\usepackage{arxiv}

\usepackage[utf8]{inputenc} 
\usepackage[T1]{fontenc}    
\usepackage{hyperref}       
\usepackage{url}            
\usepackage{booktabs}       
\usepackage{amsfonts}       
\usepackage{nicefrac}       
\usepackage{microtype}      
\usepackage{lipsum}		
\usepackage{graphicx}
\usepackage{natbib}
\usepackage{doi}
\usepackage{authblk}
\usepackage{amsmath}
\usepackage{braket}

\title{Probing Ultra-Fast Dephasing via Entangled Photon Pairs}

\date{}

\author[1,*]{Xinghua Liu}
\author[1,2,3]{Tian Li}
\author[1]{Jiaxuan Wang}
\author[1]{Mrunal R. Kamble}
\author[1]{Aleksei M. Zheltikov}
\author[1,2]{Girish S. Agarwal}
\affil[1]{Institute for Quantum Science and Engineering, Department of Physics and Astronomy, Texas A\&M University, College Station, Texas 77843-4242, USA}
\affil[2]{Department Department of Biological and Agricultural Engineering, Texas A\&M University, College Station, TX 77843, USA}
\affil[3]{(Current address) Departments of Chemistry and Physics, The University of Tennessee, Chattanooga, TN 37403, USA}
\affil[*]{xinghua@tamu.edu}

\hypersetup{
pdftitle={A template for the arxiv style},
pdfsubject={q-bio.NC, q-bio.QM},
pdfauthor={David S.~Hippocampus, Elias D.~Striatum},
pdfkeywords={First keyword, Second keyword, More},
}

\begin{document}
\maketitle

\begin{abstract}
We demonstrate how the Hong-Ou-Mandel (HOM) interference with polarization-entangled photons can be used to probe ultrafast dephasing. We can infer the optical properties like the real and imaginary parts of the complex susceptibility of the medium from changes in the position and the shape of the HOM dip. From the shift of the HOM dip, we are able to measure 22 fs dephasing time using a continuous-wave (CW) laser even with optical loss > 97~\%, while the HOM dip visibility is maintained at 92.3~\% (which can be as high as 96.7 ~\%). The experimental observations, which are explained in terms of a rigorous theoretical model, demonstrate the utility of HOM interference in probing ultrafast dephasing. 
\end{abstract}

\section{Introduction}
The Hong-Ou-Mandel interference\cite{PhysRevLett.59.2044, Bouchard_2020, RevModPhys.94.025007} was a turning point in the field of optical interference. It opened up several new pathways- it established the fact that interference between two independent but identical photons is possible, leading to the remarkable possibility of detecting time delays at the femtosecond scale and even lower\cite{doi:10.1126/sciadv.aap9416, Volkovich:20}. The original HOM has been generalized in many different ways, including the observation of HOM in frequency domain\cite{PhysRevLett.124.143601, PhysRevLett.126.123601, PhysRevX.10.031031, doi:10.1073/pnas.2010827117}. Recently, many different applications of HOM interference have appeared \cite{doi:10.1063/1.4724105,yang2019two}. It has now become a standard tool to establish the indistinguishable nature of photons \cite{Santori2002, PhysRevResearch.4.013037}. The nonidentical photons can also result in HOM dip but with reduced visibility depending on the nature of photons \cite{PhysRevLett.126.063602,Wiegner_2011,PhysRevLett.123.080401, doi:10.1126/sciadv.abm8171, PhysRevLett.129.093604}. The HOM with quantum dot sources\cite{PhysRevLett.107.157402, PhysRevLett.126.063602} and plasmonic systems\cite{PhysRevApplied.1.034004, PhysRevApplied.2.014004, DuttaGupta:14} have been studied. Applications of HOM in the context of quantum microscopy and quantum imaging have been considered\cite{Ndagano2022, doi:10.1126/sciadv.abj2155, defienne2022pixel}. A significant application of HOM is to produce NOON state, which is of particular importance in quantum metrology \cite{doi:10.1080/00107510802091298, PhysRevA.104.062613} and quantum microscopy \cite{Ono2013}. Recently several papers have discussed the possibility of determining the absorptive and dispersive properties of matter by using HOM \cite{kalashnikov2017quantum, doi:10.1021/jacs.1c02514, dorfman2021hong}. The paper by Eshun ~\textit{et al.} \cite{doi:10.1021/jacs.1c02514} successfully used organic samples through their HOM dip visibilities were only around 40~\% due to wide spectral ranges of pulsed lasers. 

In this paper, we show how HOM interference can be a useful probe of optical absorption and dispersion at the ultra-fast scale. We use a CW pump to produce polarization-entangled photons and change the polarization of the signal photon to study the HOM with identical photons. One of the two photons is sent to a medium, which is effectively characterized by a frequency-dependent optical susceptibility. The results show that the frequency-dependence of the complex susceptibility breaks the indistinguishability of the two single photons leading to changes in the interference dip. These changes are in the HOM dip location, visibility, and symmetry. Using a CW laser, the tunable dephasing time can be measured down to 22 fs or more. Our HOM dip visibility can be as high as 96.7~\% and maintains 92.3~\% even for an open system with optical loss > 97~\%. Moreover, we found that both the absorption and dispersion of a sample can be inferred from the tilt and shift of the HOM dip, which provides another useful approach for characterizing a sample's optical properties at the ultra-fast scale. The experimental observations are fully explained in terms of a rigorous theoretical model.

\section{Experiment}
\begin{figure}[htbp]
  \centering
 
  \includegraphics[width=10.3cm]{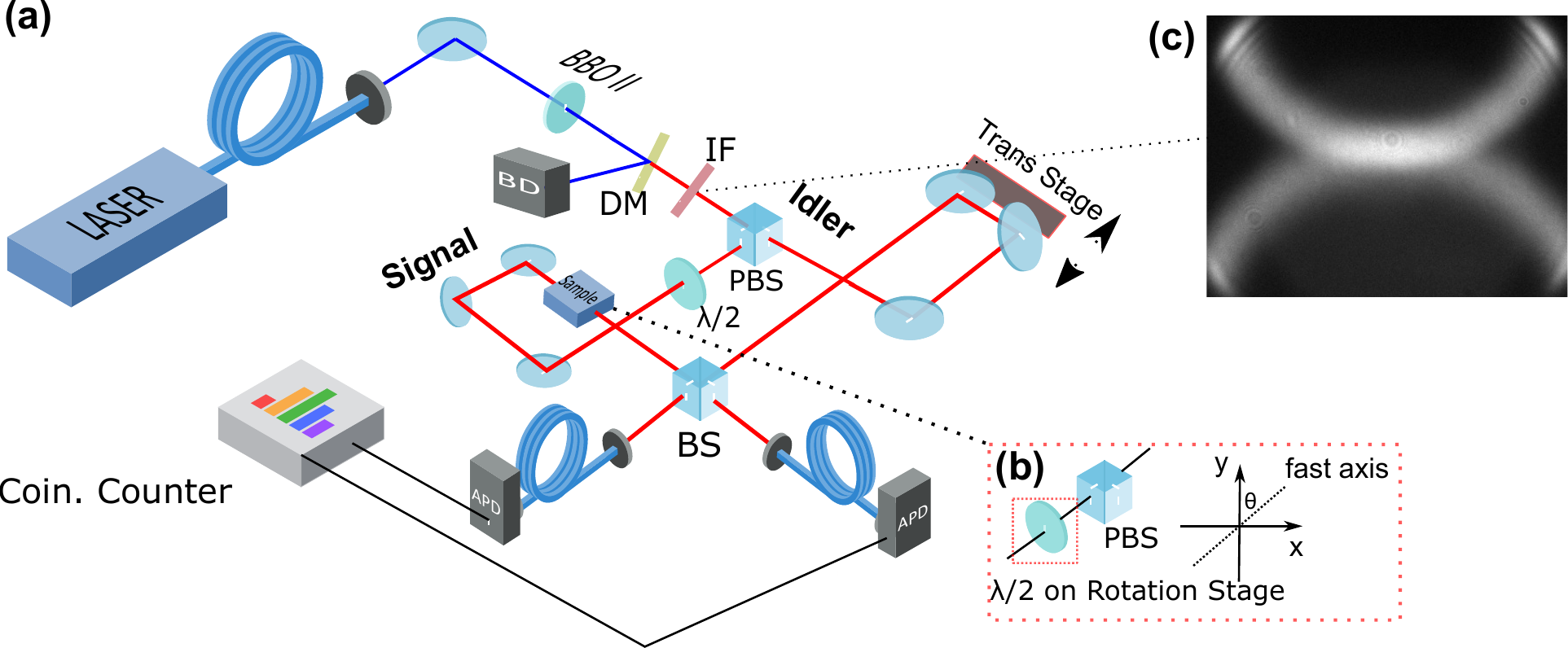}
\caption{(a) Experimental setup comprised of an SPDC and a HOM interferometer. BBO: $\beta$-Barium bOrate crystal cut for type-II phase matching for 404~nm, DM: long-pass dichroic mirror, IF: interference filter, PBS: polarizing beam splitter, BS: 50:50 non-polarizing beam splitter, APD: avalanche photodiode for single-photon counting. (b) A testing sample is mimicked by a PBS in conjunction with a Multi-Order (MO) \textit{or} a Zero-Order (ZO) half-wave plate (HWP) mounted on a motorized rotational stage. (c) Image of the SPDC rings captured by an EMCCD camera.}
 \label{fig:setup}
\end{figure}

Our experimental setup is shown in Fig~ \ref{fig:setup}(a), where an entangled photon pair generation by spontaneous parametric down-conversion (SPDC) and a HOM interferometer comprises the bulk of it. A fiber-coupled 404~nm CW laser (Toptica iBeam Smart) is used to pump a 1~mm-thick $\beta$-Barium bOrate (BBO) crystal cut for type-II phase-matching. The down-converted photon pairs (signal and idler) are frequency degenerate ($\omega_s = \omega_i = \frac{1}{2} \omega_p$) and are orthogonally polarized. The BBO crystal is mounted on a 3-axis rotational stage to satisfy the co-linear phase-matching condition, where the generated signal and idler rings are tangent. The phase-matching condition is verified by an Electron-Multiplying CCD (EMCCD) camera (Andor iXon 897), as shown in Fig.~\ref{fig:setup}(c).
A 505~nm long-pass dichroic mirror (DM) and an 810-10~nm interference filter (IF) are used to block residual pump photons and to select the entangled photon pairs.

The polarization-entangled photon pairs are then separated by a polarizing beam splitter (PBS) so that the signal and idler photons can be in different optical paths. After the PBS, the idler photons transverse an optical delay stage composed of a retro-reflecting hollow roof prism (Thorlab HR1015-AG, ultrafast-enhanced silver coating) mounted on a motorized translational stage (Thorlab PT1-Z8). In this configuration, the relative optical delay between the photon pairs can be adjusted by increasing or decreasing the path length of the idler photons while keeping their propagation direction unchanged. The motorized stage has a resolution of 29~nm, corresponding to a time delay resolution of 0.2~fs. In the optical path of signal photons, a zero-order (ZO) half-wave plate (HWP) is used to match their polarization to that of the idler photons. The signal photons then pass a `mock sample' composed of an HWP followed by a PBS (Fig.~\ref{fig:setup}.(b)), and are combined with the idler photons on a 50:50 non-polarizing beam splitter (BS) to form a HOM interferometer. The outputs of the BS are coupled through two single-mode fibers to two single-photon avalanche photodiodes (Excelitas, SPCM-AQRH-13-FC). Single photon counting and coincidence countings rates are acquired by a coincidence counter (IDQ, ID900) within a 4~ns sampling window. 
 
To achieve the photon pair indistinguishability required by the HOM interferometer, the signal and idler photons must be overlapped perfectly in space. Due to the SPDC conversion ratio for BBO crystal being in the order of $10^{-11}$, it's unsuitable to use the entangled photon pairs themselves to perform alignment for the interferometer. Therefore, an additional 780~nm CW external cavity diode laser ( not shown in Fig~\ref{fig:setup}(a))) is used for the alignment purpose. We again use the EMCCD camera to verify that this `guiding' laser is aligned with the down-converted photons both in the near and far fields. With the aid of the guiding laser, the perfectness of overlap between signal and idler photons can be judged by interference fringes having good visibilities. After this interference alignment, the ECDL is then replaced by the 404~nm diode laser. Since the coherence length of the  ECDL is much longer than the down-converted photons, additional pre-scanning is required to find the 0-delay position, where the optical paths of the signal and idler photons are identical.

\begin{figure}[h!]
\centering\includegraphics[width=9cm]{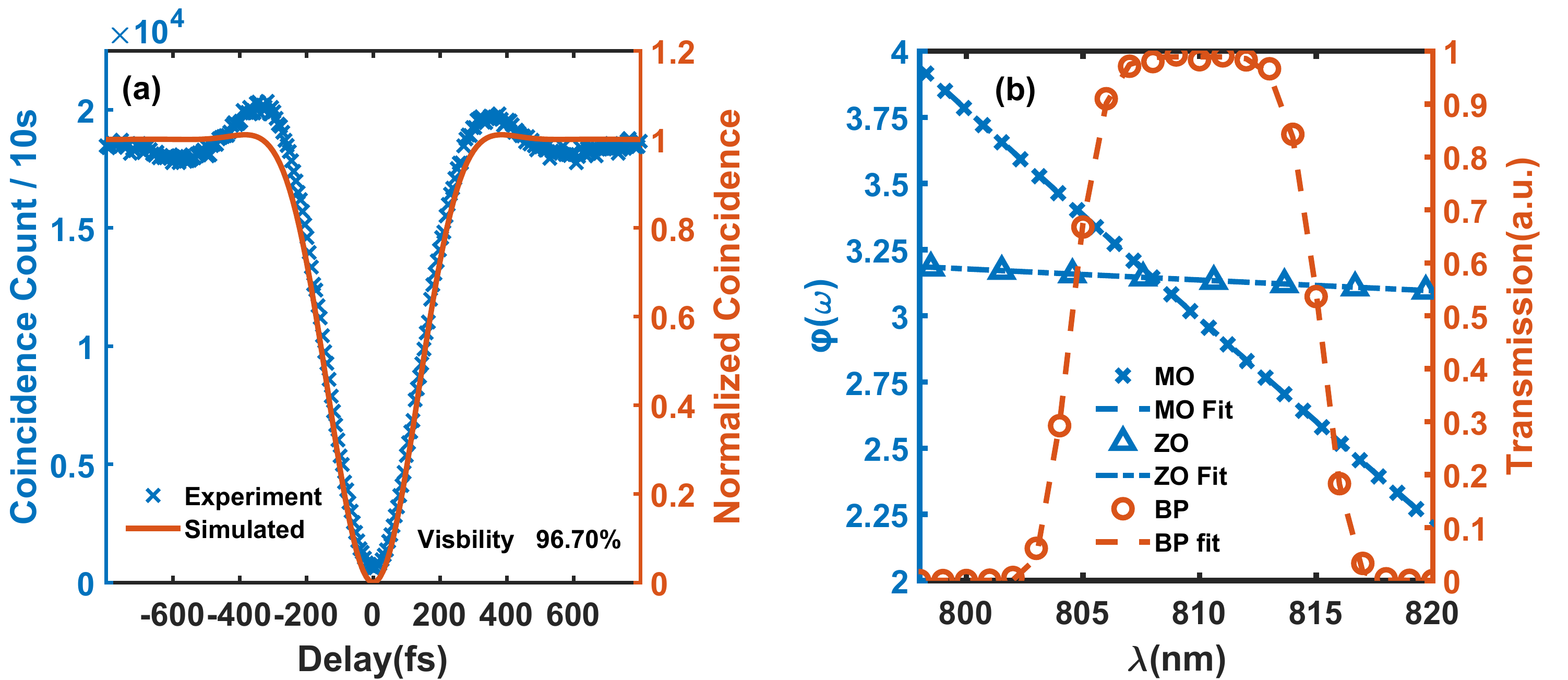}
\caption{(a) Experimental (blue cross, scale on the left) and simulated (red line, normalized, with the scale on the right) coincidence counts without a sample. (b) Experimental and linear-fitted phase delay $\phi(\omega)$ for the ZO HWP (triangle, blue with scale on the left) and the MO HWP (blue circle, with the scale on the left). Transmission data (red circle, with the scale on the right) and super-Gaussian fit (red dash line,  with the scale on the right) for the band-pass (BP) filter. Data was acquired from the Thorlab specification~\cite{Thorlab1, Thorlab2}.}
\label{fig:waveplate}
\end{figure}

In this experiment, two types of HWPs at 808~nm are used in conjunction with PBS to mimic the transmission property of a sample. 
One type of HWP is a Multi-Order (MO) HWP, and the other is a Zero-Order (ZO) HWP. The MO HWP is made of crystalline quartz, a type of birefringent material that has different refraction indexes along two orthogonal axes (ordinary axis refraction index $n_o(\omega)$ and extraordinary axis refraction index $n_e(\omega)$). 
 \begin{equation}
    \Delta\Phi = \frac{2\pi L(n_o(\omega) - n_e(\omega))}{\lambda},
 \end{equation}
where $\Delta \Phi$ is the phase delay between the ordinary and extraordinary axes, and L is the length of the wave plate. For an MO HWP at 808~nm, L is selected to maintain $\Delta \Phi = \pi$ \textit{only} for photons with a wavelength of 808~nm. Therefore, the MO HWP would introduce different phase delays for different wavelengths, as shown in Fig.~\ref{fig:waveplate}(b). On the other hand, the ZO HWP, which is composed of two MO HWPs, can offer less wavelength dependency for the phase delay, and is this considered a "perfect" HWP in this experiment.

\section{Results and Analysis}

The typical bandwidth of entangled photon pairs produced by a type-II BBO crystal is ~20~nm \cite{Kim:05}. In our experiment, it's determined by the 810-10~nm premium hard-coated band-pass filter (Thorlabs, FBH810-10), whose transmission curve is shown in Fig.~\ref{fig:waveplate}(b) (red circles, with the scale on the right). The band-pass filter also blocks residual pump photons. Furthermore, the filter applies an additional spectral selection for the entangled photons to reduce photon pair distinguishability and thus increase the HOM dip depth \cite{PhysRevA.57.R2289}. We can fit the transmission function of the band-pass filter with a super-Gaussian function, $T_f(\omega) = a e^{-(\frac{\omega-b}{c})^6} + d$, as shown in Fig.~\ref{fig:waveplate}(b)(red dash line, with the scale on the right).
When the pump power is $50$~mW, and the coincidence window is $4$~ns, we are able to acquire $1.84\times10^{4}$ photon pairs per 10~s. The visibility V of a HOM dip is defined as: 
\begin{equation}
    V = \frac{R_c(\infty) - R_c(0)}{R_c(\infty) + R_c(0)},
    \label{V}
\end{equation}   
where $R_c(\infty)$ is the coincidence counts rate far from the coherence region, and $R_c(0)$ is the coincidence counts rate for the zero delay between the signal and idler photons (detailed derivation of $R_c$ is in the next section). Our HOM dip's visibility is measured to be ($96.7 \pm 0.1\%$) as shown in Fig.~\ref{fig:waveplate}(a).

\begin{figure}[htbp]
\centering\includegraphics[width=9cm]{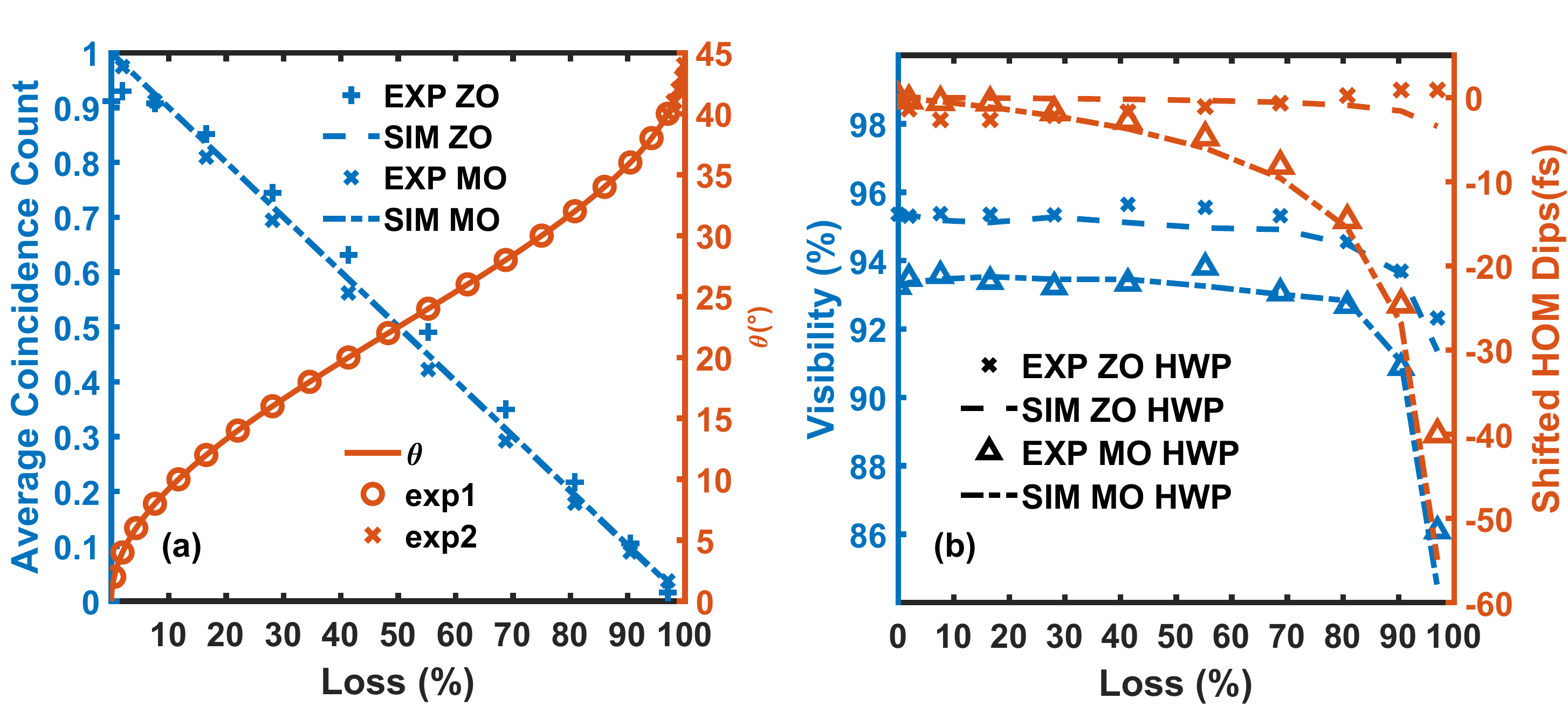}
\caption{Experimental and simulation results for (a) Change of average coincidence counts versus optical loss (blue, with the scale on the left axis).  Optical loss as a function of $\theta$ (red, with the scale on the right axis). (b) Experimental and simulation results for visibility versus optical loss (blue, with the scale on the left axis). Shifted position of HOM dip versus optical loss (red, with the scale on the right axis), the dephasing time $T_2$ can be extracted from this information (Fig.~\ref{fig:lorentz} (b)). Optical loss is calculated based on a `perfect' HWP as discussed in Eqn.~\ref{eqn:loss}.}
\label{fig:shift}
\end{figure}

To achieve a \textit{continuously adjustable} absorption and dispersion, we put the MO/ZO HWP on a rotational stage, so that the angle $\theta$ between the fast axis of the HWP and the polarization of the signal photon can be turned according to a complex transmission function $T(\omega)$:
\begin{equation}
T(\omega)=\text{cos}\frac{\phi(\omega)}{2}-i\text{sin}\frac{\phi(\omega)}{2}\text{cos}2\theta,
\end{equation}
where $\phi(\omega)$ is the phase delay between the fast and slow axes, as shown in Fig.~\ref{fig:setup}(b) (detailed derivation of this equation can be found in the next section). For a `perfect' HWP, the phase delay should be independent of wavelength, i.e., $\phi(\omega) = \pi$. In this ideal case, the transmission function reduces to $T(\omega) = -i\text{cos}2\theta$ and $|T(\omega)|^2 = \text{cos}^2(2\theta)$. Therefore, loss for 'perfect' HWP in our system will be 
\begin{eqnarray}
Loss = 1-\text{cos}^2(2\theta)
\label{eqn:loss}
\end{eqnarray}
The corresponding loss function with respect to $\theta$ for this `perfect' HWP is shown in  Fig.~\ref{fig:shift}(a) (red curve).

\begin{figure}[htbp]
\centering\includegraphics[width=9cm]{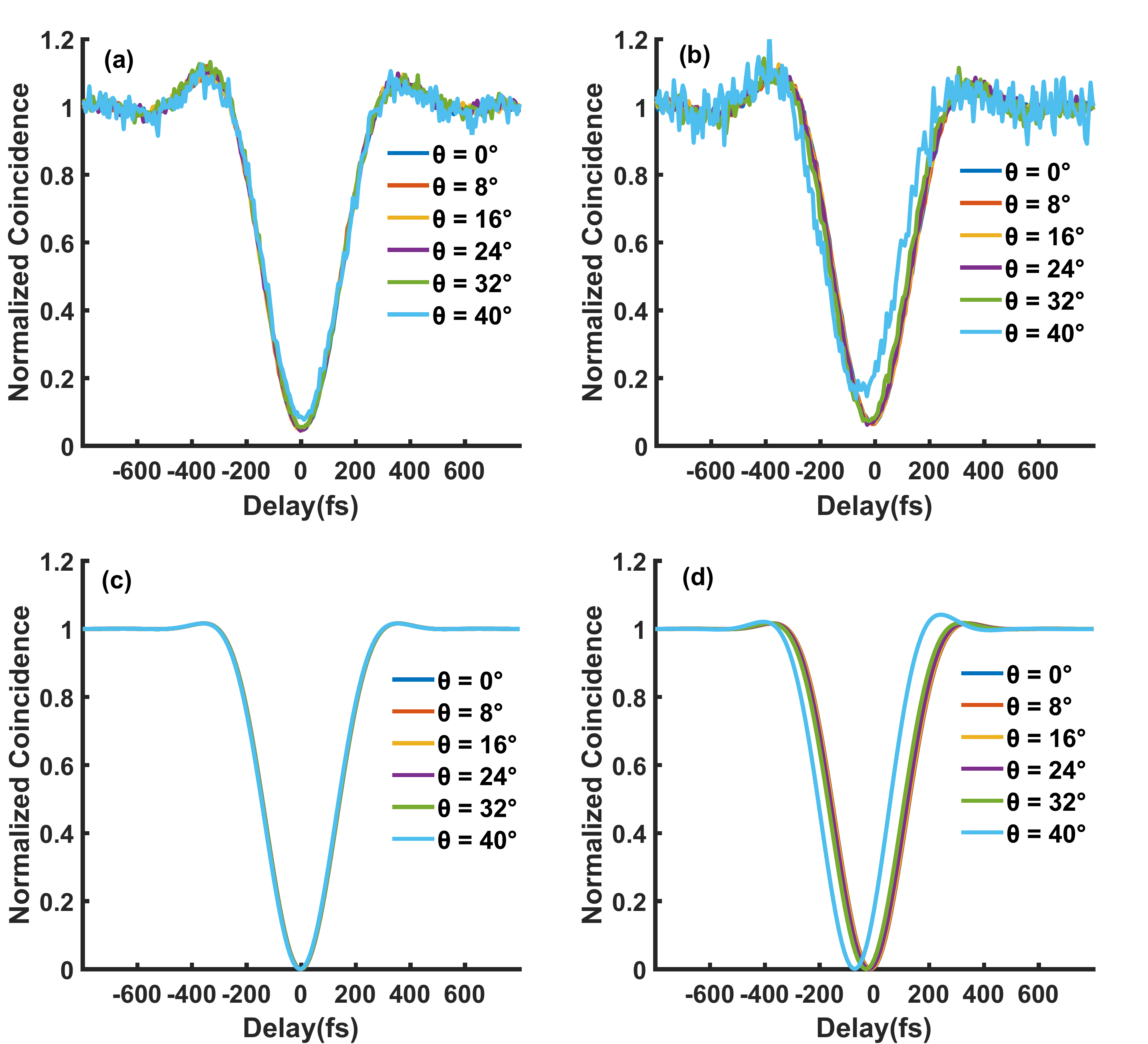}
\caption{Experimental and simulation results for optical loss less than $97\%$ ($\theta = 0^{\circ},8^{\circ},16^{\circ},24^{\circ},32^{\circ}, 40^{\circ}$). (a) Experimental results for the ZO HWP. (b) Experimental results for the MO HWP. (c) Simulation results for the ZO HWP. (d) Simulation results for the MO HWP. The simulation results are based on the Theoretical Model, Eqn.~\ref{eqn:int}. }
\label{fig:exp1}
\end{figure}

\begin{figure}[htbp]
\centering\includegraphics[width=9cm]{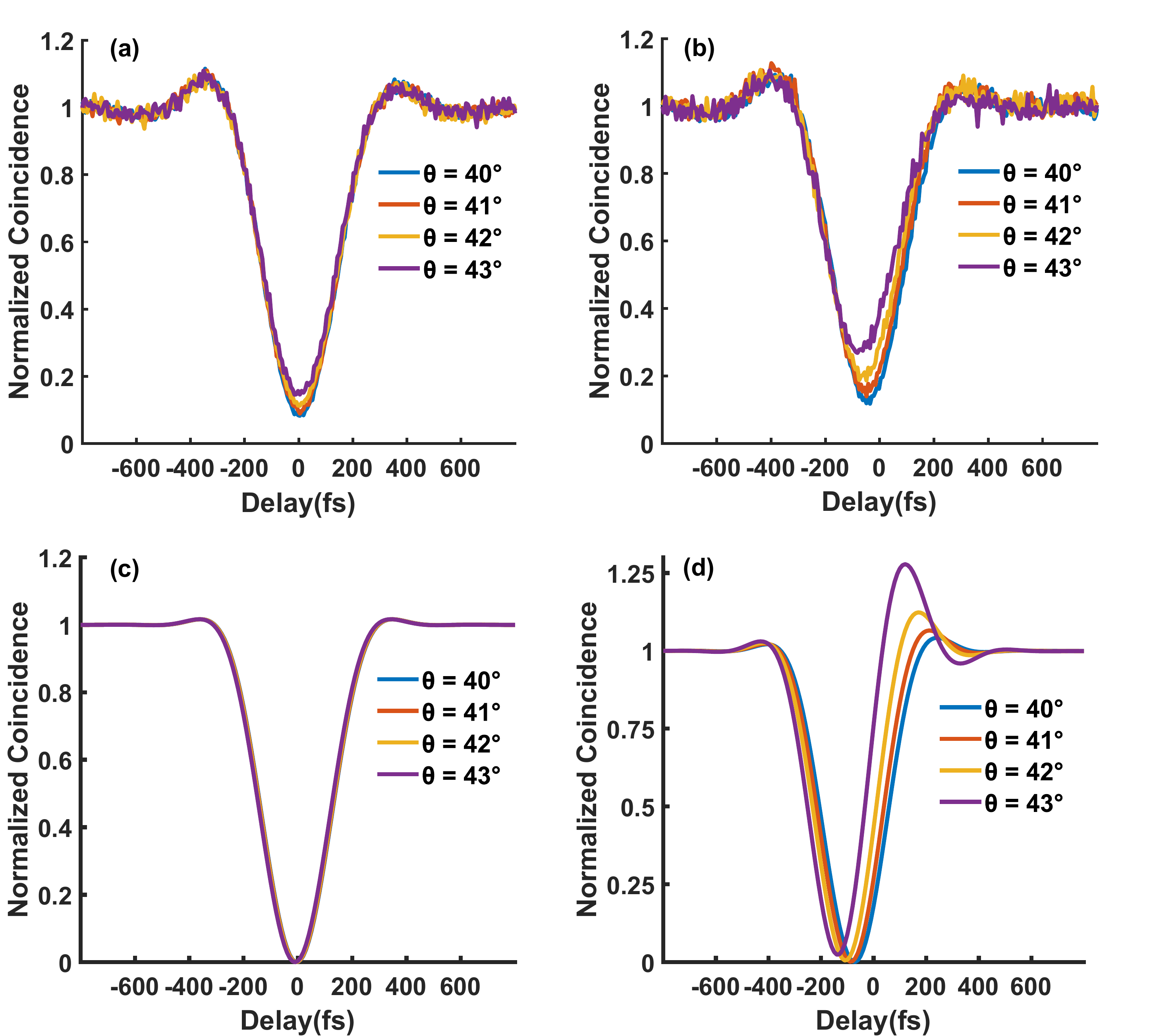}
\caption{Experimental and simulation results for optical loss greater than $97\%$ ($\theta = 40^{\circ},41^{\circ},42^{\circ},43^{\circ}$). (a) Experimental results for the ZO HWP. (b) Experimental results for the MO half-wave. (c) Simulation results for the ZO HWP. (d) Simulation results for the MO HWP. The simulation results are based on the Theoretical Model, Eqn.~\ref{eqn:int}.}
\label{fig:exp2}
\end{figure}

Figures~\ref{fig:exp1} and~\ref{fig:exp2} are the experimental and simulation results for the normalized HOM dips, respectively. The dips are normalized to the average background counts in the uncorrelated region, i.e., $R_c(\infty)$ in Eq.~\ref{V}. The data are collected within two regimes: 1) $0^{\circ} \leq \theta \leq  40^{\circ}$ and 2) $40^{\circ} \leq \theta \leq  45^{\circ}$, which are separated by whether the optical loss is greater than $97\%$. Figure~\ref{fig:exp1} is for the first regime where the angle changes every $8^{\circ}$ and acquisition time is 10~s per data point. The left-hand-side panels are the experimental results (Fig.~\ref{fig:exp1}(a)) and simulation results (Fig.~\ref{fig:exp1}(c)) for the ZO HWP, while the right-hand-side panels are the experimental results (Fig.~\ref{fig:exp1}(b)) and simulation results (Fig.~\ref{fig:exp1}(d)) for the MO HWP. We can see that in the case of ZO HWP, all HOM dips overlap pretty well, which implies wavelength-independent loss (no dispersion dependence) has no effect on the indistinguishability of the photon pairs. Whereas in the case of MO HWP, the minima of these HOM dips shift to the $-\tau$ direction as $\theta$ increases. 


The experimental and simulation results for the second regime where $\theta \geq 40^\circ$ are shown in Fig.~\ref{fig:exp2}. The HOM dips now were taken with only $1^{\circ}$ increment, i.e., $\theta =  40^{\circ},41^{\circ},42^{\circ},43^{\circ}$, and acquisition time is 50~s per data point. Same as Fig.~\ref{fig:exp1}, the left- and right-hand side panels are the experimental (Fig.~\ref{fig:exp2}(a) and~(b))and simulation (Fig.~\ref{fig:exp2}(c) and~(d)) results for the ZO and MO HWPs respectively. We can still see that in the case of the ZO HWP, all HOM dips overlap except that their depths reduce as the angle (i.e., absorption) increases, which can be attributed to the fact that the artificial coincidence now takes a larger portion of the total coincidence. The artificial coincidence $R_{AC}$ is defined as the average coincidence counts within a time window when both detectors register fully random (uncorrelated) events. In our case, $R_{AC} \sim (I_s +I_i)^2\sim (I_s + \bar t I_{i0})^2 $, where $I_s, I_i$ are the signal and idler photon numbers, $I_{i0}$ is the photon number in the idler beam path if there is no sample present, $\bar t$ is the average transmission rate of the sample. Since the coincidence counts from correlated photons are proportional to the product of signal and idler photon numbers ($R_{c} \sim I_s \times I_i \sim \bar t I_s I_{i0})$, thus the ratio of an artificial coincidence to real coincidence ($R_{AC}/R_c  \sim \frac{1 + 2 \bar t}{\bar t}$) approaches $\infty$ as $\bar t \rightarrow 0$. In the case of the MO HWP, in addition to the locations of these HOM dips shift to the $-\tau$ direction as in Fig.~\ref{fig:exp1}(b)(d), shapes of these HOM dips are not symmetric anymore, which implies a change of spectral mode of the signal photons.     

To further investigate the difference between the wavelength-independent and the wavelength-dependent transmission, we plot the average coincidence counts $R_c(\infty)$ change as a function of optical loss introduced by a 'perfect' HWP in Fig.~\ref{fig:shift}(a) blue curve. We also plot the optical loss as a function of $\theta$ in Fig.~\ref{fig:shift}(a) red curve. 
The measured average coincidence counts are normalized by the coincidence counts in the incoherent regime $R_c(\infty)$ where there is no loss. We can see that the average coincidence counts maintain a linear behavior with respect to the optical loss, which implies that both ZO and MO HWPs act like a photon number attenuator as their spectral modes now is an irrelevant point of concern. However, their spectral modes are modified with large dispersion, and the photon pair indistinguishability is subsequently reduced.

Moreover, we plot the change in visibility and shift of HOM dip versus optical loss in Fig.~\ref{fig:shift}(b). It contains the simulated and experimental results for visibility changes of MO HWP and ZO HWP, respectively (the blue curve, with the scale on the left axis). Again, for small absorption, visibilities remain unchanged for both ZO and MO HWPs, whereas for large absorption, visibilities for wavelength-independent transmission (ZO HWP) degrade significantly. This observation can still be attributed to the effect of aforementioned artificial coincidence. Both experimental and theoretical result shows that the visibility could reach 95.7~\% for zero loss and remain 92.7~\% at the optical loss of 97~\% for wavelength-independent transmission (ZO HWP). This observation can still be attributed to the effect of the aforementioned artificial coincidence. The visibility starts at 93.5~\% at zero loss and drops to 86.1~\% with 97~\% optical loss for wavelength-dependent transmission (MO HWP). We can also see that visibilities deteriorate more aggressively for wavelength-dependent transmission (using MO HWP) due to the combined impact of both the artificial coincidence and distortion of the HOM dip.

This effect can also be seen in the change of HOM dip position versus optical loss (Fig.~\ref{fig:shift}(b) red curve, with the scale on the right axis). For wavelength-independent transmission (using ZO HWP), the center positions of the HOM dips remain unchanged when optical loss increases. In contrast, for wavelength-dependent transmission (using MO HWP), the HOM dips hardly shift with small dispersion ($\delta\tau < 5$~fs when the loss is less than $40~\%$), yet with large dispersion, they shift $\sim~40$~fs towards $-\tau$ direction. The shift and distortion of HOM dip caused by wavelength-dependent transmission (using MO HWP) could be modeled as the effect of a fictitious oscillator with ultra-fast dephasing (Eqn.~\ref{eqn:Tlorentz}).
\begin{eqnarray}
    T(\omega) 
        &=& \text{exp}({\frac{iBL/c}{\Omega- \omega - i/T_2}})
\end{eqnarray}
where $B$ is the Bouguer coefficient, $L$ is the length of the sample, c is the speed of light, $\Omega$ is the resonant frequency, and $T_2$ is the dephasing time. The detailed derivation is shown in the next section. From this model, we can extract the ultra-fast dephasing time for the fictitious oscillator as shown in Fig.~\ref{fig:lorentz}(b).

\section{Theoretical Model}
The entangled twin-photon state generated by SPDC can be written as
\begin{eqnarray}
\label{eqn:1}
\ket{\Phi} = \int^{\infty}_{-\infty} \int^{\infty}_{-\infty} d\omega_s d\omega_i f(\omega_s, \omega_i) a^{\dagger}_s(\omega_s)a^{\dagger}_i(\omega_i)\ket{0_s,0_i},
\end{eqnarray} where $a^{\dagger}_s$ ($a^{\dagger}_i$) is the creation operator for the signal (idler) photon, and $f(\omega_s, \omega_i)$ is the joint spectral amplitude (JSA) of the entangled photons, which depends both on the spectrum of the pump and the proprieties (length, cutting angle and phase-matching condition) of the BBO crystal. A polarizing beam splitter separates the twin photons. In one of the paths, the signal photon travels through the sample of interest 
\begin{eqnarray}
a^{'}_s(\omega) = T(\omega) a_s(\omega) + \eta(\omega)a_v(\omega),
\end{eqnarray}
 where $T(\omega)$ is the transmission function that carries the sample's optical properties to the photon field; $a_v(\omega)$ is the vacuum noise operator, and $\eta(\omega)=\sqrt{1-|T(\omega)|^2}$. $a^{'}_s(\omega)$ and $a^{'\dagger}_s(\omega^{'})$ satisfy the commutation relations $[a^{'}_s(\omega),a^{'\dagger}_s(\omega^{'}) ] = \delta(\omega - \omega^{'})$. Note that the vacuum noise term does not contribute to the correlation because it is normal-ordered. $T(\omega)$ can also be written in terms of sample length $L$, speed of light $c$, and complex refractive index $n+i\kappa$ of the sample as 
\begin{eqnarray}
T(\omega) = e^{i\omega(n(\omega) + i\kappa(\omega))L/c}.
\label{eqn:3}
\end{eqnarray} 
Meanwhile, the idler photon in the other path acquires a phase delay $\tau$ introduced by a retro-reflecting mirror mounted on a translational stage
\begin{eqnarray}
a^{'}_i(\omega) = a_i(\omega)e^{-i\omega\tau}.
\end{eqnarray}
The photons are then recombined on a 50:50 beam splitter and produce the outputs modes ($c(\omega)$, $d(\omega)$)
\begin{eqnarray}
\begin{pmatrix} 
c(\omega) \\
d(\omega)
\end{pmatrix}
= \frac{1}{\sqrt(2)}
\begin{pmatrix} 
1 & i \\
i & 1
\end{pmatrix}
\begin{pmatrix} 
a^{'}_s(\omega) \\
a^{'}_i(\omega)
\end{pmatrix},
\end{eqnarray}
$c(\omega)$, $d(\omega)$ are then collected by two SPADs and analyzed by the coincidence counter. The detector response time $\tau_d$ is in the order of $1\;ns$ while the correlation time of SPDC photons is usually $0.1-1 \; ps$ \cite{PhysRevA.79.063846}. Thus, we can assume the detector response time is much longer than the correlation time of the entangled photons, then the averaged coincidence rate ($R_c(\tau)$) is given by
\begin{eqnarray}
R_c(\tau) = \int^{\infty}_{-\infty} \int^{\infty}_{-\infty} d t d t^{'} \langle d^{\dagger}(t)c^{\dagger}(t + t^{'})c(t + t^{'})d(t)\rangle
\end{eqnarray}
Changing the modes to frequency domain $d(t)=\frac{1}{\sqrt{2\pi}}\int d\omega_{1}d(\omega_{1})e^{-i\omega_{1}t}$ and $c(t+t^{'})=\frac{1}{\sqrt{2\pi}}\int d\omega_{2}c(\omega_{2})e^{-i\omega_{2}(t+t^{'})}$, we obtain
\begin{eqnarray}
R_{c}(\tau)&=&\frac{1}{8\pi^2}\int^{\infty}_{-\infty} \int^{\infty}_{-\infty} d\omega_{1}d\omega_{2}f^{*}(\omega_{1},\omega_{2})f(\omega_{1},\omega_{2})T^{*}(\omega_{1})T(\omega_{1})\nonumber    \\
&-&\frac{1}{8\pi^2}\int^{\infty}_{-\infty} \int^{\infty}_{-\infty} d\omega_{1}d\omega_{2}f^{*}(\omega_{1},\omega_{2})f(\omega_{2},\omega_{1})T^{*}(\omega_{1})T(\omega_{2})e^{-i(\omega_{1}-\omega_{2})\tau}.
\label{eqn:int1}
\end{eqnarray}
With super-Gaussian fit for band-pass filter $T_f(\omega) = a e^{-(\frac{\omega-b}{c})^6} + d$, Eqn.~\ref{eqn:int1} is modified to 

\begin{eqnarray}
R_{c}(\tau)&=&\frac{1}{8\pi^2}\int^{\infty}_{-\infty} \int^{\infty}_{-\infty} d\omega_{1}d\omega_{2}f^{*}(\omega_{1},\omega_{2})f(\omega_{1},\omega_{2})T^{*}(\omega_{1})T(\omega_{1})|T_f(\omega_1)|^2 |T_f(\omega_2)|^2 \nonumber    \\
&-&\frac{1}{8\pi^2}\int^{\infty}_{-\infty} \int^{\infty}_{-\infty} d\omega_{1}d\omega_{2}f^{*}(\omega_{1},\omega_{2})f(\omega_{2},\omega_{1})T^{*}(\omega_{1})T(\omega_{2}) \nonumber\\
& &|T_f(\omega_1)|^2 |T_f(\omega_2)|^2e^{-i(\omega_{1}-\omega_{2})\tau}.
\label{eqn:int}
\end{eqnarray}
It is not possible to obtain analytical result now and therefore all simulations are done numerically using Eq.~\ref{eqn:int}.

In this study, we consider a Gaussian entangled two-photon distribution with both photons centered at half of the pump frequency  $\omega_{p}/2$
\begin{equation}
f(\omega_{1},\omega_{2})=\frac{1}{\sqrt{2\pi\sigma_{P}\sigma_{-}}}e^{-(\omega_{1}+\omega_{2}-\omega_{p})^{2}/(16\sigma_{P}^{2})}e^{-(\omega_{1}-\omega_{2})^{2}/(4\sigma_{-}^{2})},
\end{equation}
where $\omega_p$ and $\sigma_p$ are the center frequency and bandwidth of the pump laser. Note that $f(\omega_{1},\omega_{2})$ is normalized by $\int\int|f(\omega_{1},\omega_{2})|^2 d\omega_{1} d \omega_{2} = 1$. For the $405\;nm$ CW laser in this experiment, we have $\omega_p = 3.069\; eV$ and $\sigma_p = 0.149\;meV$. $\sigma_-$ ($6.8\;meV$) is the bandwidth of photon pairs, where $\sigma_- \sim \frac{1}{T_e}$. $T_e$ is the entanglement time of the photon pairs and obtained from the width of the HOM dip (Fig. \ref{fig:waveplate}.(a)).   $\omega_1, \omega_2$ are frequencies for signal and idler photon, respectively. The center frequencies of signal photon and idler photon are $\omega_1 = \omega_2 = \frac{\omega_p}{2} = 1.535 \; eV$. 
\begin{figure}[htbp]
\centering\includegraphics[width=9cm]{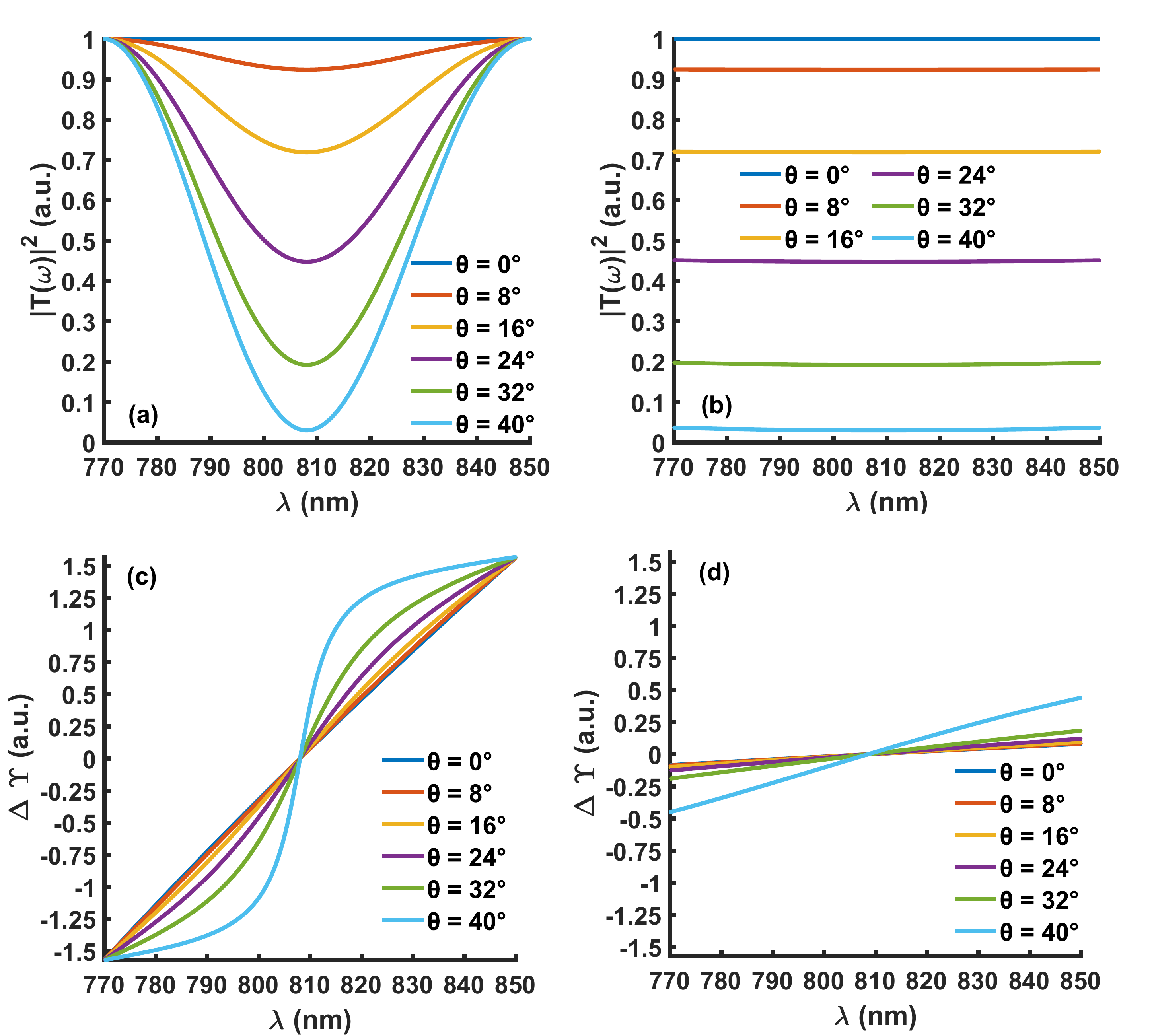}
\caption{(a)Transmission spectrum ($|T(\omega)|^2$) for MO HWP and PBS. (b) Transmission spectrum ($|T(\omega)|^2$) for ZO HWP and PBS. (c)Phase change spectrum ($\Delta\Upsilon$) for MO HWP and PBS. (d)Phase change spectrum ($\Delta\Upsilon$) for ZO HWP and PBS.  }
\label{fig:absopr}
\end{figure}

The test sample in this experiment consists of an HWP on a rotational stage and a PBS (Fig. \ref{fig:setup}.(b)). The HWP transforms the orthogonal components of a light state as \cite{agarwal2012quantum}
\begin{equation}
\left(\begin{array}{c}
H_{out}\\
V_{out}
\end{array}\right)=J\,'\left(\omega\right)\left(\begin{array}{c}
H_{in}\\
V_{in}
\end{array}\right).
\end{equation}
The transformation matrix ($J^{'}(\omega)$) is the rotation transformation of the Jones matrix for a wave plate, where
\begin{equation}
J\,'\left(\omega\right)=R^{-1}\left(\theta\right)J\left(\omega\right)R\left(\theta\right).
\end{equation}
The Jones matrix ($J(\omega)$) for a wave plate can be written as,
\begin{equation}
J\left(\omega\right)=\left(\begin{array}{cc}
e^{i\frac{\phi\left(\omega\right)}{2}} & 0\\
0 & e^{-i\frac{\phi\left(\omega\right)}{2}}
\end{array}\right),
\end{equation}
where $\phi(\omega)$ is the phase delay between the wave plate's fast and slow polarization axis. 
The rotation matrix ($R(\theta)$) can be represented as,
\begin{equation}
R\left(\theta\right)=\left(\begin{array}{cc}
\text{cos}\theta & \text{sin}\theta\\ 
-\text{sin}\theta & \text{cos}\theta
\end{array}\right).
\end{equation}
where $\theta$ is the angle between the vertical direction and fast axis of the HWP, as shown in Fig.\ref{fig:setup}(b).\\
We obtain
\begin{equation}
J\,'\left(\omega\right)=\left(\begin{array}{cc}
\text{cos}\frac{\phi\left(\omega\right)}{2}+i\text{sin}\frac{\phi\left(\omega\right)}{2}\text{cos}2\theta & i\text{sin}\frac{\phi\left(\omega\right)}{2}\text{sin}2\theta\\
i\text{sin}\frac{\phi\left(\omega\right)}{2}\text{sin}2\theta & \text{cos}\frac{\phi\left(\omega\right)}{2}-i\text{sin}\frac{\phi\left(\omega\right)}{2}\text{cos}2\theta
\end{array}\right).
\end{equation}
For an input state $\left|V_{in}\right\rangle$ with only the vertical polarization and output state $\left|V_{out}\right\rangle$ selected by PBS to maintain only vertical polarization component,
\begin{equation}
\left|V_{out}\right\rangle =[\text{cos}\frac{\phi\left(\omega\right)}{2}-i\text{sin}\frac{\phi\left(\omega\right)}{2}\text{cos}2\theta]\left|V_{in}\right\rangle .
\end{equation}
Thus, the transmission function can be written as
\begin{equation}
T(\omega)=\text{cos}\frac{\phi(\omega)}{2}-i\text{sin}\frac{\phi(\omega)}{2}\text{cos}2\theta:=|T(\omega)|e^{i\Delta\Upsilon(\omega)}.
\label{eqn:Tsys}
\end{equation}
Where $\Delta\Upsilon(\omega) = \text{tan}^{-1}[-\text{tan}\frac{\phi(\omega)}{2}\text{cos}2\theta]$. 
$|T(\omega)|^2$ corresponds to transmission spectrum and the phase $\Delta\Upsilon(\omega)$ corresponds to phase delay. They are shown in Fig.\ref{fig:absopr} as functions of wavelength for $\theta = 40^\circ$.\\

In the frequency range of interest, $\phi(\omega)$ can be approximately treated as a linear function $\phi(\omega)=\alpha\omega + \beta$  as shown in Fig.~ \ref{fig:waveplate}(b). $\alpha$ is $40.30\; eV^{-1}$ for a MO HWP and $2.18\; eV^{-1}$ for a ZO HWP. $\beta$ is $-60.23$ for a MO HWP and $-0.21$ for a ZO HWP. Carrying out the double integral Eqn.~(\ref{eqn:int1}) (note that the super-Gaussian fit $T_f(\omega)$ is not applied here), we obtain
\begin{eqnarray}
R_{c}(\tau)&=&\frac{1}{8\pi^2}\{1+\frac{\text{cos}2\theta}{2}e^{-\frac{\sigma_{-}^{2}(\alpha^{2}+4\tau^{2})}{8}}[\text{sinh}(\sigma_{-}^{2}\alpha\tau)-\text{cosh}(\sigma_{-}^{2}\alpha\tau)\text{cos}2\theta]\nonumber    \\
&-&\frac{\text{sin}^{2}2\theta}{2}[e^{-\frac{1}{2}\sigma_{P}^{2}\alpha^{2}}(1-e^{-\frac{1}{8}\sigma_{-}^{2}\alpha^{2}})\text{cos}(\beta+\frac{\alpha\omega_{p}}{2})+1]\}.
\label{eqn:res}
\end{eqnarray}

In this experiment, a frequency filter is placed before the PBS to get rid of the influence of the pump laser and can be approximated by a super-Gaussian function, as shown in Fig.~\ref{fig:waveplate} (c).

The linear susceptibility for a damped harmonic oscillator with the eigenfrequency  resonant at $\Omega$ can be modeled with a Lorentz function\cite{stenzel2022light}:
\begin{equation}
    \chi(\omega)=\frac{q^2}{m}\frac{1}{-\omega^2-\frac{2i\omega}{T_2} + \Omega^2}
\end{equation}
Where $\Omega$ is the resonance frequency, $T_2$ is the dephasing time, $q$ is the charge $m$ is the mass. For a near-resonant absorption $\omega \sim \Omega$ ($\Omega$ = 808~nm while $\omega_s$ range from 803~nm to 813~nm), we simplified the above equation to 
\begin{equation}
    \chi(\omega)=\frac{B}{\Omega-\omega-i/T_2}
\end{equation}
Where $B = \frac{q^2}{2m\Omega}$. From Eqn.~(\ref{eqn:3}), the corresponding  transmission function:
\begin{eqnarray}
     T(\omega) = \text{exp}({\frac{iBL/c}{\Omega- \omega - i/T_2}})
    \label{eqn:Tlorentz}
\end{eqnarray}

\begin{figure}[h!]
\centering\includegraphics[width=9cm]{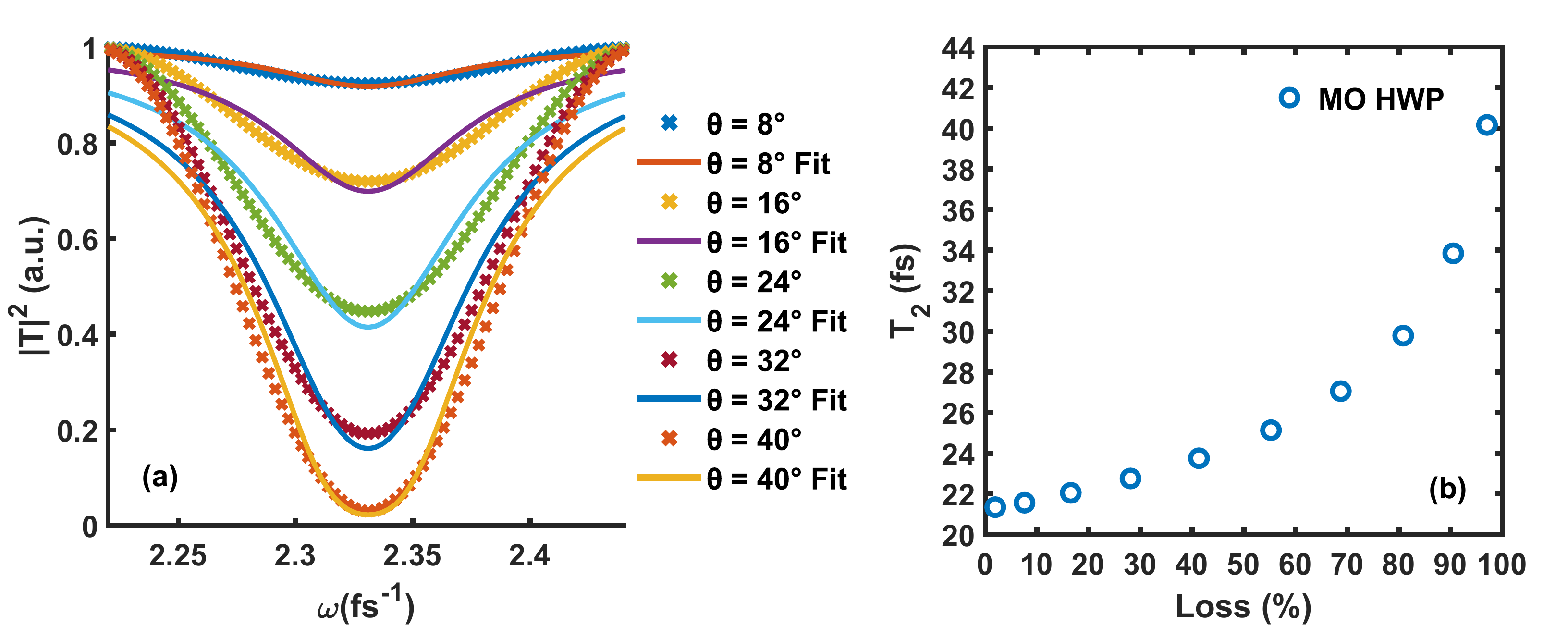}
\caption{(a) Comparison of transmission spectrum $|T(\omega)|^2$ of the system (Eqn.~\ref{eqn:Tsys}) and Lorentz fit (Eqn.~\ref{eqn:Tlorentz}) for MO HWP and PBS orientated at different angles. (b) Dephasing time ($T_2$) versus optical loss based on the Lorentz model for a two-level system. Optical loss is calculated based on a `perfect' HWP as discussed in Eqn.~\ref{eqn:loss}.}
\label{fig:lorentz}
\end{figure}

 Comparison of transmission spectrum $|T(\omega)|^2$ of system (Eqn.~\ref{eqn:Tsys}) versus Lorentz fit (Eqn.~\ref{eqn:Tlorentz}) is shown on Fig.~\ref{fig:lorentz}(a). From the formula, one can obtain the dephasing time $T_2$ of the sample versus $\theta$. The calculated dephasing time $T_2$ for a MO HWP and PBS is shown in Fig.~\ref{fig:lorentz}(b), which shows the capacity of measuring $22\;fs$ dephasing time with the experimental scheme.

\section{Conclusion}
We demonstrate an experimental scheme and present a theoretical framework to study the effects of tunable optical properties of a sample on photon indistinguishability in a HOM interferometer. We continuously change a mock sample's transmission/dispersion by varying the orientation of a ZO/MO HWP followed by a PBS. Our results indicate that the structures of a HOM dip (dip location, visibility, and symmetry) would contain information about a sample's optical properties. Specifically, a wavelength-dependent transmission/dispersion would partially break the indistinguishability and thus change the dip location, visibility, and symmetry of the HOM dip, as is also predicted by our theoretical model. Moreover, a 22~fs dephasing time is obtained using our experimental scheme even with optical loss $> 97\%$ (while HOM dip visibility is maintained at $92.3~\%$), which provides a useful approach for characterizing a sample's optical properties even in an enormously lossy environment.

\section*{Funding}

This research was supported in part by the Robert A. Welch Foundation (Grant A-1801-20180324, A-1943-20210327) and the Air Force Office of
Scientific Research (Award No. FA-9550-20-1-0366).

\section*{Disclosures}
\noindent The authors declare no conflicts of interest.

\section*{Data Availability}

Data underlying the results presented in this paper are not publicly available at this time but may be obtained from the authors upon reasonable request.

\bibliographystyle{unsrtnat}
\bibliography{references}  

\begin{thebibliography}{38}
\providecommand{\natexlab}[1]{#1}
\providecommand{\url}[1]{\texttt{#1}}
\expandafter\ifx\csname urlstyle\endcsname\relax
  \providecommand{\doi}[1]{doi: #1}\else
  \providecommand{\doi}{doi: \begingroup \urlstyle{rm}\Url}\fi

\bibitem[Hong et~al.(1987)Hong, Ou, and Mandel]{PhysRevLett.59.2044}
C.~K. Hong, Z.~Y. Ou, and L.~Mandel.
\newblock Measurement of subpicosecond time intervals between two photons by
  interference.
\newblock \emph{Phys. Rev. Lett.}, 59:\penalty0 2044--2046, Nov 1987.
\newblock \doi{10.1103/PhysRevLett.59.2044}.
\newblock URL \url{https://link.aps.org/doi/10.1103/PhysRevLett.59.2044}.

\bibitem[Bouchard et~al.(2020)Bouchard, Sit, Zhang, Fickler, Miatto, Yao,
  Sciarrino, and Karimi]{Bouchard_2020}
Fr{\'{e}}d{\'{e}}ric Bouchard, Alicia Sit, Yingwen Zhang, Robert Fickler,
  Filippo~M Miatto, Yuan Yao, Fabio Sciarrino, and Ebrahim Karimi.
\newblock Two-photon interference: the {Hong-Ou-Mandel} effect.
\newblock \emph{Reports on Progress in Physics}, 84\penalty0 (1):\penalty0
  012402, dec 2020.
\newblock \doi{10.1088/1361-6633/abcd7a}.
\newblock URL \url{https://doi.org/10.1088/1361-6633/abcd7a}.

\bibitem[Hochrainer et~al.(2022)Hochrainer, Lahiri, Erhard, Krenn, and
  Zeilinger]{RevModPhys.94.025007}
Armin Hochrainer, Mayukh Lahiri, Manuel Erhard, Mario Krenn, and Anton
  Zeilinger.
\newblock Quantum indistinguishability by path identity and with undetected
  photons.
\newblock \emph{Rev. Mod. Phys.}, 94:\penalty0 025007, Jun 2022.
\newblock \doi{10.1103/RevModPhys.94.025007}.
\newblock URL \url{https://link.aps.org/doi/10.1103/RevModPhys.94.025007}.

\bibitem[Lyons et~al.(2018)Lyons, Knee, Bolduc, Roger, Leach, Gauger, and
  Faccio]{doi:10.1126/sciadv.aap9416}
Ashley Lyons, George~C. Knee, Eliot Bolduc, Thomas Roger, Jonathan Leach,
  Erik~M. Gauger, and Daniele Faccio.
\newblock Attosecond-resolution {Hong-Ou-Mandel} interferometry.
\newblock \emph{Science Advances}, 4\penalty0 (5):\penalty0 eaap9416, 2018.
\newblock \doi{10.1126/sciadv.aap9416}.
\newblock URL \url{https://www.science.org/doi/abs/10.1126/sciadv.aap9416}.

\bibitem[Volkovich and Shwartz(2020)]{Volkovich:20}
Sergey Volkovich and Sharon Shwartz.
\newblock Subattosecond {X-ray} {Hong--Ou--Mandel} metrology.
\newblock \emph{Opt. Lett.}, 45\penalty0 (10):\penalty0 2728--2731, May 2020.
\newblock \doi{10.1364/OL.382044}.
\newblock URL \url{https://opg.optica.org/ol/abstract.cfm?URI=ol-45-10-2728}.

\bibitem[Joshi et~al.(2020)Joshi, Farsi, Dutt, Kim, Ji, Zhao, Bishop, Lipson,
  and Gaeta]{PhysRevLett.124.143601}
Chaitali Joshi, Alessandro Farsi, Avik Dutt, Bok~Young Kim, Xingchen Ji, Yun
  Zhao, Andrew~M. Bishop, Michal Lipson, and Alexander~L. Gaeta.
\newblock Frequency-domain quantum interference with correlated photons from an
  integrated microresonator.
\newblock \emph{Phys. Rev. Lett.}, 124:\penalty0 143601, Apr 2020.
\newblock \doi{10.1103/PhysRevLett.124.143601}.
\newblock URL \url{https://link.aps.org/doi/10.1103/PhysRevLett.124.143601}.

\bibitem[Hiekkam\"aki and Fickler(2021)]{PhysRevLett.126.123601}
Markus Hiekkam\"aki and Robert Fickler.
\newblock High-dimensional two-photon interference effects in spatial modes.
\newblock \emph{Phys. Rev. Lett.}, 126:\penalty0 123601, Mar 2021.
\newblock \doi{10.1103/PhysRevLett.126.123601}.
\newblock URL \url{https://link.aps.org/doi/10.1103/PhysRevLett.126.123601}.

\bibitem[Devaux et~al.(2020)Devaux, Mosset, Moreau, and
  Lantz]{PhysRevX.10.031031}
Fabrice Devaux, Alexis Mosset, Paul-Antoine Moreau, and Eric Lantz.
\newblock Imaging spatiotemporal {Hong-Ou-Mandel} interference of biphoton
  states of extremely high schmidt number.
\newblock \emph{Phys. Rev. X}, 10:\penalty0 031031, Aug 2020.
\newblock \doi{10.1103/PhysRevX.10.031031}.
\newblock URL \url{https://link.aps.org/doi/10.1103/PhysRevX.10.031031}.

\bibitem[Cerf and Jabbour(2020)]{doi:10.1073/pnas.2010827117}
Nicolas~J. Cerf and Michael~G. Jabbour.
\newblock Two-boson quantum interference in time.
\newblock \emph{Proceedings of the National Academy of Sciences}, 117\penalty0
  (52):\penalty0 33107--33116, 2020.
\newblock \doi{10.1073/pnas.2010827117}.
\newblock URL \url{https://www.pnas.org/doi/abs/10.1073/pnas.2010827117}.

\bibitem[Crespi et~al.(2012)Crespi, Lobino, Matthews, Politi, Neal, Ramponi,
  Osellame, and O’Brien]{doi:10.1063/1.4724105}
Andrea Crespi, Mirko Lobino, Jonathan C.~F. Matthews, Alberto Politi, Chris~R.
  Neal, Roberta Ramponi, Roberto Osellame, and Jeremy~L. O’Brien.
\newblock Measuring protein concentration with entangled photons.
\newblock \emph{Applied Physics Letters}, 100\penalty0 (23):\penalty0 233704,
  2012.
\newblock \doi{10.1063/1.4724105}.
\newblock URL \url{https://doi.org/10.1063/1.4724105}.

\bibitem[Yang et~al.(2019)Yang, Xu, and Giovannetti]{yang2019two}
Yu~Yang, Luping Xu, and Vittorio Giovannetti.
\newblock Two-parameter {Hong-Ou-Mandel} dip.
\newblock \emph{Scientific Reports}, 9\penalty0 (1):\penalty0 1--16, 2019.

\bibitem[Santori et~al.(2002)Santori, Fattal, Vu{\v{c}}kovi{\'{c}}, Solomon,
  and Yamamoto]{Santori2002}
Charles Santori, David Fattal, Jelena Vu{\v{c}}kovi{\'{c}}, Glenn~S. Solomon,
  and Yoshihisa Yamamoto.
\newblock Indistinguishable photons from a single-photon device.
\newblock \emph{Nature}, 419\penalty0 (6907):\penalty0 594--597, Oct 2002.
\newblock ISSN 1476-4687.
\newblock \doi{10.1038/nature01086}.
\newblock URL \url{https://doi.org/10.1038/nature01086}.

\bibitem[Schofield et~al.(2022)Schofield, Clear, Hoggarth, Major, McCutcheon,
  and Clark]{PhysRevResearch.4.013037}
Ross~C. Schofield, Chloe Clear, Rowan~A. Hoggarth, Kyle~D. Major, Dara P.~S.
  McCutcheon, and Alex~S. Clark.
\newblock Photon indistinguishability measurements under pulsed and continuous
  excitation.
\newblock \emph{Phys. Rev. Research}, 4:\penalty0 013037, Jan 2022.
\newblock \doi{10.1103/PhysRevResearch.4.013037}.
\newblock URL \url{https://link.aps.org/doi/10.1103/PhysRevResearch.4.013037}.

\bibitem[Ollivier et~al.(2021)Ollivier, Thomas, Wein, de~Buy~Wenniger, Coste,
  Loredo, Somaschi, Harouri, Lemaitre, Sagnes, Lanco, Simon, Anton, Krebs, and
  Senellart]{PhysRevLett.126.063602}
H.~Ollivier, S.~E. Thomas, S.~C. Wein, I.~Maillette de~Buy~Wenniger, N.~Coste,
  J.~C. Loredo, N.~Somaschi, A.~Harouri, A.~Lemaitre, I.~Sagnes, L.~Lanco,
  C.~Simon, C.~Anton, O.~Krebs, and P.~Senellart.
\newblock {Hong-Ou-Mandel} interference with imperfect single photon sources.
\newblock \emph{Phys. Rev. Lett.}, 126:\penalty0 063602, Feb 2021.
\newblock \doi{10.1103/PhysRevLett.126.063602}.
\newblock URL \url{https://link.aps.org/doi/10.1103/PhysRevLett.126.063602}.

\bibitem[Wiegner et~al.(2011)Wiegner, von Zanthier, and Agarwal]{Wiegner_2011}
R~Wiegner, J~von Zanthier, and G~S Agarwal.
\newblock Quantum interference and non-locality of independent photons from
  disparate sources.
\newblock \emph{Journal of Physics B: Atomic, Molecular and Optical Physics},
  44\penalty0 (5):\penalty0 055501, feb 2011.
\newblock \doi{10.1088/0953-4075/44/5/055501}.
\newblock URL \url{https://doi.org/10.1088/0953-4075/44/5/055501}.

\bibitem[Deng et~al.(2019)Deng, Wang, Ding, Duan, Qin, Chen, He, He, Li, Li,
  Peng, Matekole, Byrnes, Schneider, Kamp, Wang, Dowling, H\"ofling, Lu,
  Scully, and Pan]{PhysRevLett.123.080401}
Yu-Hao Deng, Hui Wang, Xing Ding, Z.-C. Duan, Jian Qin, M.-C. Chen, Yu~He,
  Yu-Ming He, Jin-Peng Li, Yu-Huai Li, Li-Chao Peng, E.~S. Matekole, Tim
  Byrnes, C.~Schneider, M.~Kamp, Da-Wei Wang, Jonathan~P. Dowling, Sven
  H\"ofling, Chao-Yang Lu, Marlan~O. Scully, and Jian-Wei Pan.
\newblock Quantum interference between light sources separated by 150 million
  kilometers.
\newblock \emph{Phys. Rev. Lett.}, 123:\penalty0 080401, Aug 2019.
\newblock \doi{10.1103/PhysRevLett.123.080401}.
\newblock URL \url{https://link.aps.org/doi/10.1103/PhysRevLett.123.080401}.

\bibitem[Koong et~al.(2022)Koong, Cygorek, Scerri, Santana, Park, Song, Gauger,
  and Gerardot]{doi:10.1126/sciadv.abm8171}
Zhe~Xian Koong, Moritz Cygorek, Eleanor Scerri, Ted~S. Santana, Suk~In Park,
  Jin~Dong Song, Erik~M. Gauger, and Brian~D. Gerardot.
\newblock Coherence in cooperative photon emission from indistinguishable
  quantum emitters.
\newblock \emph{Science Advances}, 8\penalty0 (11):\penalty0 eabm8171, 2022.
\newblock \doi{10.1126/sciadv.abm8171}.
\newblock URL \url{https://www.science.org/doi/abs/10.1126/sciadv.abm8171}.

\bibitem[Su et~al.(2022)Su, Zhong, Zhang, Li, Zou, Wang, Yan, and
  Zhu]{PhysRevLett.129.093604}
Keyu Su, Yi~Zhong, Shanchao Zhang, Jianfeng Li, Chang-Ling Zou, Yunfei Wang,
  Hui Yan, and Shi-Liang Zhu.
\newblock Quantum interference between nonidentical single particles.
\newblock \emph{Phys. Rev. Lett.}, 129:\penalty0 093604, Aug 2022.
\newblock \doi{10.1103/PhysRevLett.129.093604}.
\newblock URL \url{https://link.aps.org/doi/10.1103/PhysRevLett.129.093604}.

\bibitem[Polyakov et~al.(2011)Polyakov, Muller, Flagg, Ling, Borjemscaia,
  Van~Keuren, Migdall, and Solomon]{PhysRevLett.107.157402}
Sergey~V. Polyakov, Andreas Muller, Edward~B. Flagg, Alex Ling, Natalia
  Borjemscaia, Edward Van~Keuren, Alan Migdall, and Glenn~S. Solomon.
\newblock Coalescence of single photons emitted by disparate single-photon
  sources: The example of inas quantum dots and parametric down-conversion
  sources.
\newblock \emph{Phys. Rev. Lett.}, 107:\penalty0 157402, Oct 2011.
\newblock \doi{10.1103/PhysRevLett.107.157402}.
\newblock URL \url{https://link.aps.org/doi/10.1103/PhysRevLett.107.157402}.

\bibitem[Di~Martino et~al.(2014)Di~Martino, Sonnefraud, Tame, K\'ena-Cohen,
  Dieleman, \"Ozdemir, Kim, and Maier]{PhysRevApplied.1.034004}
G.~Di~Martino, Y.~Sonnefraud, M.~S. Tame, S.~K\'ena-Cohen, F.~Dieleman,
  \ifmmode \mbox{\c{S}}\else \c{S}\fi{}.~K. \"Ozdemir, M.~S. Kim, and S.~A.
  Maier.
\newblock Observation of quantum interference in the plasmonic {Hong-Ou-Mandel}
  effect.
\newblock \emph{Phys. Rev. Applied}, 1:\penalty0 034004, Apr 2014.
\newblock \doi{10.1103/PhysRevApplied.1.034004}.
\newblock URL \url{https://link.aps.org/doi/10.1103/PhysRevApplied.1.034004}.

\bibitem[Cai et~al.(2014)Cai, Li, Ren, Zou, Xiong, Lei, Liu, Guo, and
  Guo]{PhysRevApplied.2.014004}
Yong-Jing Cai, Ming Li, Xi-Feng Ren, Chang-Ling Zou, Xiao Xiong, Hua-Lin Lei,
  Bi-Heng Liu, Guo-Ping Guo, and Guang-Can Guo.
\newblock High-visibility on-chip quantum interference of single surface
  plasmons.
\newblock \emph{Phys. Rev. Applied}, 2:\penalty0 014004, Jul 2014.
\newblock \doi{10.1103/PhysRevApplied.2.014004}.
\newblock URL \url{https://link.aps.org/doi/10.1103/PhysRevApplied.2.014004}.

\bibitem[Gupta and Agarwal(2014)]{DuttaGupta:14}
S.~Dutta Gupta and G.~S. Agarwal.
\newblock Two-photon quantum interference in plasmonics: theory and
  applications.
\newblock \emph{Opt. Lett.}, 39\penalty0 (2):\penalty0 390--393, Jan 2014.
\newblock \doi{10.1364/OL.39.000390}.
\newblock URL \url{https://opg.optica.org/ol/abstract.cfm?URI=ol-39-2-390}.

\bibitem[Ndagano et~al.(2022)Ndagano, Defienne, Branford, Shah, Lyons,
  Westerberg, Gauger, and Faccio]{Ndagano2022}
Bienvenu Ndagano, Hugo Defienne, Dominic Branford, Yash~D. Shah, Ashley Lyons,
  Niclas Westerberg, Erik~M. Gauger, and Daniele Faccio.
\newblock Quantum microscopy based on {Hong-Ou-Mandel} interference.
\newblock \emph{Nature Photonics}, 16\penalty0 (5):\penalty0 384--389, May
  2022.
\newblock ISSN 1749-4893.
\newblock \doi{10.1038/s41566-022-00980-6}.
\newblock URL \url{https://doi.org/10.1038/s41566-022-00980-6}.

\bibitem[Camphausen et~al.(2021)Camphausen, Álvaro Cuevas, Duempelmann,
  Terborg, Wajs, Tisa, Ruggeri, Cusini, Steinlechner, and
  Pruneri]{doi:10.1126/sciadv.abj2155}
Robin Camphausen, Álvaro Cuevas, Luc Duempelmann, Roland~A. Terborg, Ewelina
  Wajs, Simone Tisa, Alessandro Ruggeri, Iris Cusini, Fabian Steinlechner, and
  Valerio Pruneri.
\newblock A quantum-enhanced wide-field phase imager.
\newblock \emph{Science Advances}, 7\penalty0 (47):\penalty0 eabj2155, 2021.
\newblock \doi{10.1126/sciadv.abj2155}.
\newblock URL \url{https://www.science.org/doi/abs/10.1126/sciadv.abj2155}.

\bibitem[Defienne et~al.(2022)Defienne, Cameron, Ndagano, Lyons, Reichert,
  Zhao, Harvey, Charbon, Fleischer, and Faccio]{defienne2022pixel}
Hugo Defienne, Patrick Cameron, Bienvenu Ndagano, Ashley Lyons, Matthew
  Reichert, Jiuxuan Zhao, Andrew~R Harvey, Edoardo Charbon, Jason~W Fleischer,
  and Daniele Faccio.
\newblock Pixel super-resolution with spatially entangled photons.
\newblock \emph{Nature communications}, 13\penalty0 (1):\penalty0 1--9, 2022.

\bibitem[Dowling(2008)]{doi:10.1080/00107510802091298}
Jonathan~P. Dowling.
\newblock Quantum optical metrology – the lowdown on high-{N00N} states.
\newblock \emph{Contemporary Physics}, 49\penalty0 (2):\penalty0 125--143,
  2008.
\newblock \doi{10.1080/00107510802091298}.
\newblock URL \url{https://doi.org/10.1080/00107510802091298}.

\bibitem[Wang and Agarwal(2021)]{PhysRevA.104.062613}
J.~Wang and G.~S. Agarwal.
\newblock Quantum fisher information bounds on precision limits of circular
  dichroism.
\newblock \emph{Phys. Rev. A}, 104:\penalty0 062613, Dec 2021.
\newblock \doi{10.1103/PhysRevA.104.062613}.
\newblock URL \url{https://link.aps.org/doi/10.1103/PhysRevA.104.062613}.

\bibitem[Ono et~al.(2013)Ono, Okamoto, and Takeuchi]{Ono2013}
Takafumi Ono, Ryo Okamoto, and Shigeki Takeuchi.
\newblock An entanglement-enhanced microscope.
\newblock \emph{Nature Communications}, 4\penalty0 (1):\penalty0 2426, Sep
  2013.
\newblock ISSN 2041-1723.
\newblock \doi{10.1038/ncomms3426}.
\newblock URL \url{https://doi.org/10.1038/ncomms3426}.

\bibitem[Kalashnikov et~al.(2017)Kalashnikov, Melik-Gaykazyan, Kalachev, Yu,
  Kuznetsov, and Krivitsky]{kalashnikov2017quantum}
Dmitry~A Kalashnikov, Elizaveta~V Melik-Gaykazyan, Alexey~A Kalachev, Ye~Feng
  Yu, Arseniy~I Kuznetsov, and Leonid~A Krivitsky.
\newblock Quantum interference in the presence of a resonant medium.
\newblock \emph{Scientific reports}, 7\penalty0 (1):\penalty0 1--8, 2017.

\bibitem[Eshun et~al.(2021)Eshun, Gu, Varnavski, Asban, Dorfman, Mukamel, and
  Goodson]{doi:10.1021/jacs.1c02514}
Audrey Eshun, Bing Gu, Oleg Varnavski, Shahaf Asban, Konstantin~E. Dorfman,
  Shaul Mukamel, and Theodore Goodson.
\newblock Investigations of molecular optical properties using quantum light
  and {Hong–Ou–Mandel} interferometry.
\newblock \emph{Journal of the American Chemical Society}, 143\penalty0
  (24):\penalty0 9070--9081, 2021.
\newblock \doi{10.1021/jacs.1c02514}.
\newblock URL \url{https://doi.org/10.1021/jacs.1c02514}.
\newblock PMID: 34124903.

\bibitem[Dorfman et~al.(2021)Dorfman, Asban, Gu, and Mukamel]{dorfman2021hong}
Konstantin~E Dorfman, Shahaf Asban, Bing Gu, and Shaul Mukamel.
\newblock {Hong-Ou-Mandel} interferometry and spectroscopy using entangled
  photons.
\newblock \emph{Communications Physics}, 4\penalty0 (1):\penalty0 1--7, 2021.

\bibitem[{Thorlab}({\natexlab{a}})]{Thorlab1}
{Thorlab}.
\newblock {Data for 808 Multi-Order Waveplates}.
\newblock
  \url{https://www.thorlabs.com/images/popupimages/808_WPData_Multi_xls.xls},
  {\natexlab{a}}.

\bibitem[{Thorlab}({\natexlab{b}})]{Thorlab2}
{Thorlab}.
\newblock {Data for 810/10nm Bandpass filter}.
\newblock \url{https://www.thorlabs.com/images/tabimages/FBH810-10.xlsx},
  {\natexlab{b}}.

\bibitem[Kim and Grice(2005)]{Kim:05}
Yoon-Ho Kim and Warren~P. Grice.
\newblock Measurement of the spectral properties of the two-photon state
  generated via type ii spontaneous parametric downconversion.
\newblock \emph{Opt. Lett.}, 30\penalty0 (8):\penalty0 908--910, Apr 2005.
\newblock \doi{10.1364/OL.30.000908}.
\newblock URL \url{http://opg.optica.org/ol/abstract.cfm?URI=ol-30-8-908}.

\bibitem[Grice et~al.(1998)Grice, Erdmann, Walmsley, and
  Branning]{PhysRevA.57.R2289}
W.~P. Grice, R.~Erdmann, I.~A. Walmsley, and D.~Branning.
\newblock Spectral distinguishability in ultrafast parametric down-conversion.
\newblock \emph{Phys. Rev. A}, 57:\penalty0 R2289--R2292, Apr 1998.
\newblock \doi{10.1103/PhysRevA.57.R2289}.
\newblock URL \url{https://link.aps.org/doi/10.1103/PhysRevA.57.R2289}.

\bibitem[Blauensteiner et~al.(2009)Blauensteiner, Herbauts, Bettelli, Poppe,
  and H\"ubel]{PhysRevA.79.063846}
Bibiane Blauensteiner, Isabelle Herbauts, Stefano Bettelli, Andreas Poppe, and
  Hannes H\"ubel.
\newblock Photon bunching in parametric down-conversion with continuous-wave
  excitation.
\newblock \emph{Phys. Rev. A}, 79:\penalty0 063846, Jun 2009.
\newblock \doi{10.1103/PhysRevA.79.063846}.
\newblock URL \url{https://link.aps.org/doi/10.1103/PhysRevA.79.063846}.

\bibitem[Agarwal(2012)]{agarwal2012quantum}
Girish~S Agarwal.
\newblock \emph{Quantum Optics}, pages 275--278.
\newblock Cambridge University Press, 2012.

\bibitem[Stenzel(2022)]{stenzel2022light}
Olaf Stenzel.
\newblock \emph{Light-Matter Interaction: A Crash Course for Students of
  Optics, Photonics and Materials Science}.
\newblock Springer Nature, 2022.

\end{thebibliography}






\end{document}